\documentclass[adraft,copyright,creativecommons]{eptcs}
\usepackage{breakurl}             
\usepackage{underscore}           
\usepackage[shortlabels]{enumitem}
\setlist{nolistsep}  

\newcommand{\Sp}{\mathcal{S}}                         
\DeclareMathAlphabet{\mathbbm}{U}{bbm}{m}{n}          
\newcommand{\IT}{\mathbbm{T}}                         
\newcommand{\IN}{\mathbbm{N}}                         
\newcommand{\ID}{\mathbbm{D}}                         
\newcommand{\IG}{\mathbbm{G}}                         
\newtheorem{defi}{Definition}
\newenvironment{definition}[1]{\begin{defi} \rm \label{df:#1} }{\end{defi}}
\newcommand{\Sec}[1]{Section~\ref{sec:#1}}
\newcommand{\df}[1]{Definition~\ref{df:#1}}
\newcommand{\plat}[1]{\raisebox{0pt}[0pt][0pt]{$#1$}} 
\newcommand{\den}[1]{\mbox{$[\hspace{-1.6pt}[$}\,#1\,\mbox{$]\hspace{-1.6pt}]$}}  
\newcommand{\rec}[1]{\plat{			      
	\stackrel{\mbox{\tiny $/$}}
	{\raisebox{-.3ex}[.3ex]{\tiny $\backslash$}}
	\!\!#1\!\!
	\stackrel{\mbox{\tiny $\backslash$}}
	{\raisebox{-.3ex}[.3ex]{\tiny $/$}} }}
\newcommand{\dom}{{\it dom}}                          
\newcommand{\Var}{{\it Var}}                      

\def\titlerunning{An Algebraic Treatment of Recursion}
\title{\titlerunning}
\author{Rob van Glabbeek
\institute{Data61, CSIRO, Sydney, Australia}
\institute{School of Computer Science and Engineering,
University of New South Wales, Sydney, Australia}
}
\def\authorrunning{R.J. van Glabbeek}
\begin{document}
\providecommand\publicationstatus{Paper dedicated to Jan Bergstra,\\
  at the occasion of his 65$^{th}$ birthday and retirement.}
\maketitle

\begin{quote}\it
Jan Bergstra has put his mark on theoretical computer science by a consistent stream of original
ideas, controversial opinions, and novel approaches. He sometimes reorganised the arena, enabling
others to follow. I, for one, might never have entered computer science if it wasn't for Jan's
support and encouragement, and will never forget the team spirit in the early days of process
algebra in his group at CWI\@. This paper is dedicated to Jan, at the occasion of his 65$^{th}$
birthday and retirement.
\end{quote}

\begin{abstract}
I review the three principal methods to assign meaning to recursion in process algebra: the
denotational, the operational and the algebraic approach, and I extend the latter to unguarded
recursion.
\end{abstract}
\advance\textheight 13.6 pt
\advance\textheight 13.6 pt
\advance\textheight 13.6 pt

\section{Process Algebra}\label{sec:pa}

In process algebra, processes are often modelled as closed terms of
single-sorted specification languages.

\begin{definition}{signatures}
Let $\Var$ be a set of variables.
A {\em signature} is a set of pairs $(f,n)$ of a {\em function symbol} $f\notin\Var$ and an {\em arity} $n \in \IN$.
The set $\IT(\Sigma)$ of {\em terms} over a signature $\Sigma$ is generated by:
\begin{itemize}
\item $\Var \subseteq \IT(\Sigma)$,
\item if $(f,n) \in \Sigma$ and $t_1,\ldots,t_n \in \IT(\Sigma)$ then
$f(t_1,\ldots,t_n) \in \IT(\Sigma)$,
\item If $V_\Sp \subseteq \Var$, ~$\Sp:V_\Sp \rightarrow \IT(\Sigma)$ and $X\in V_\Sp$,
then $\rec{X|\Sp}\in \IT(\Sigma)$.
\end{itemize}
A function $\Sp$ as appears in the last clause is called a {\em recursive specification}.  A
recursive specification $\Sp$ is often displayed as $\{X\mathbin=\Sp_{\!X\!} \mid X \mathbin\in V_{\!\Sp\!}\}$.
An occurrence of a variable $y$ in a term $t$ is {\em free} if it does not
occur\linebreak[3] in a subterm of the form $\rec{X|\Sp\!}$ with $y \mathbin\in V_{\!\Sp\!\!}$.  A term
is {\em closed} if it contains no free occurrences of variables.
\end{definition}
The \emph{semantics} of such a language is a function $\den{ \_}:\IT(\Sigma) \rightarrow (\ID^\Var\rightarrow\ID)$.
It assigns to every term $t\in\IT(\Sigma)$ its meaning $\den{t} \in \ID^\Var\!\rightarrow\ID$.
The meaning of a closed term is a {\em value} chosen from a class of values $\ID$, called a {\em domain}.
The meaning of an open term is a \emph{$\Var\!$-ary operator} on $\ID$: a function of
type $\ID^\Var\!\rightarrow\ID$. It associates a value $\den{t}(\rho)\mathbin\in\ID$ to $t$ that
depends on the choice of a \emph{valuation} $\rho\!:\Var\rightarrow\ID$.

Sometimes, only a subset of $\IT(\Sigma)$ is given a semantics, for instance by restricting to terms
satisfying a syntactic criterion of \emph{guardedness}.

Another approach lacks the recursion construct itself, but declares a single recursive specification $\Sp:V_\Sp \rightarrow \IT(\Sigma)$
for the entire language \cite{Mi89}. A term $t$ in such a language can be seen as a the term $\rec{t|\Sp}$,
obtained from $t$ by substituting, for each $Y\in V_\Sp$, $\rec{Y|\Sp}$ for each occurrence of $Y$.
Conversely, each term in the general language of \df{signatures} can be converted into the form
$\rec{t|\Sp}$ with $t$ and $\Sp$ recursion-free.

\section{Denotational, Operational and Algebraic Semantics}\label{sec:den}

The standard (denotational) semantics assigns to each function $(f,n)\in \Sigma$
an $n$-ary operator $f_n^\ID:\ID^n\!\rightarrow\ID$. The semantics of a recursion-free expression $t$
is then given by
\begin{itemize}
\item $\den{X}(\rho) = \rho(X)$ \hspace{2.088in} for $X \in \Var$, and
\item $\den{f(t_1,\dots,t_n)}(\rho)=f_n^{\ID}(\den{t_1}(\rho),\dots,\den{t_n}(\rho))$
  \qquad for $(f,n)\in\Sigma$.
\pagebreak[3]
\end{itemize}
Three approaches appear in the literature to give semantics to recursion.

The \emph{denotational} approach \cite{BHR84} recognises $\den{\Sp}$ as having type
$\ID^{\Var\setminus V_\Sp} \!\rightarrow (\ID^{V_\Sp}\!\rightarrow\ID^{V_\Sp})$ and
defines $\den{\rec{X|\Sp}}(\rho)$ for $\rho\in\ID^{\Var\setminus V_\Sp}$
to be the $X$-component of the least fixed point of $\den{\Sp}(\rho)$.
For this least fixed point to exists, either $\ID$, equipped with a suitable preorder $\sqsubseteq$,
needs to be a complete lattice, with the operators $f^\ID$ monotonic,
or $(\ID,\sqsubseteq)$ be a c.p.o., with the $f^\ID$ continuous,
or $\ID$ be a complete metric space, with the $f^\ID$ contracting (or some variation on this theme).

The \emph{operational} approach \cite{Mi89} is based on a set of inference rules that derive a collection of
(labelled) transitions between  closed terms. The semantic domain is now the collection $\IG$
of \emph{process graphs} $(S,T,I)$, with $S$ a set of states, $T$ a set of transitions between
states, and $I\in S$ an initial state, possibly subject to some cardinality restrictions.
The operational semantics $\den{P}$ of a closed term $P$ takes $S$ to be the set of closed terms,
$I=P$, and $T$ the derivable transitions. The semantics of open terms can be dealt with by encoding
the process graphs $\rho(X)$ for $X\in \Var$ as constants in an appropriate extension of the process
algebra. This approach covers the meaning of recursion constructs too.

Let \emph{guardedness} be a criterion on recursive specifications, such that if $\Sp$ is guarded then
is has a unique solution, meaning that if $\rho_i$ for $i=1,2$ are valuations with
$\rho_1(Z)=\rho_2(Z)$ for all $Z\in \Var\setminus V_\Sp$, and 
$\rho_i(X) = \den{\Sp_X}(\rho_i)$ for all $X\in V_\Sp$,
then $\rho_1(X)=\rho_2(X)$ for all $X\in V_\Sp$.
The \emph{algebraic} approach~\cite{BK86} yields a semantics for terms with guarded recursion only,
where $\den{\rec{X|\Sp}}(\rho)$ for $\rho\in\ID^{\Var\setminus V_\Sp}$
is the $X$-component of the unique solution of $\den{\Sp}(\rho)$.

\section{Extending the Algebraic Approach to Unguarded Recursion}

In \cite{vG87} I proposed an extension of the algebraic approach to unguarded recursion.
An expression $\rec{X|\Sp}$ is seen as a kind of variable, only ranging over the solutions of $\Sp\!\!$.
Taking for example ACP \cite{BK86}, interpreted in a domain of process graphs modulo strong
bisimilarity \cite{BK86},
then $\rec{X|X=aX}$ is a case of guarded recursion and denotes a specific process, namely an
$a$-loop. On the other hand, $\rec{X|X=X}$ is an unguarded recursion, and seen a variable
ranging over all processes, just like $X$ itself. In between, $\rec{X|X=X+aX}$ is a case of
unguarded recursion, and seen as a variable ranging over all processes of the form $a^*P$.

To avoid ambiguity in deciding when two, almost identical, processes $\rec{X|\Sp}$ denote the same
variable or different ones, here I formalise this approach only for terms $\rec{t|\Sp}$ where no
further recursion occurs in $t$ or $\Sp$, thus following the second approach of \Sec{pa}.

A valuation $\rho:\Var\rightarrow\ID$ is \emph{compatible} with a recursive specification $\Sp$
iff $\rho(Y) = \den{\Sp_Y}(\rho)$ for all $Y\in V_\Sp$.
The meaning $\den{t}$ of a recursion-free term $t$ in the context of a global recursive
specification $\Sp$ is now a function into $\ID$ from the set of compatible valuations only.
It is obtained from the semantics of $t$ from \Sec{den} by restricting $\dom(\den{t})$ to the
compatible valuations.

In particular, an equation $t=u$ holds under this semantics iff
$\den{t}(\rho) = \den{u}(\rho)$ for all valuations $\rho$ compatible with $\Sp$.
Hence it is equivalent to the conditional equation
$\left(\bigwedge_{X\in V_\Sp} X = \Sp_X \right)  \Rightarrow t=u$.

The laws of process algebra remain valid in this approach, including the
congruence property for recursion: if $\den{\Sp_X}(\rho)=\den{\Sp'_X}(\rho)$ for \emph{all} valuations
$\rho$, and all $X \in V_\Sp=V_{\Sp'}$ then $\den{\rec{t|\Sp}}=\den{\rec{t|\Sp'}}$.

\bibliographystyle{eptcsini}
\bibliography{bergstra}
\end{document}